\documentclass[preprint,journal]{vgtc}                % final (journal style)
\ifpdf%                                % if we use pdflatex
  \pdfoutput=1\relax                   % create PDFs from pdfLaTeX
  \usepackage{graphicx}                % allow us to embed graphics files
  \DeclareGraphicsExtensions{.pdf,.png,.jpg,.jpeg} % for pdflatex we expect .pdf, .png, or .jpg files
\else%                                 % else we use pure latex
  \usepackage{graphicx}                % allow us to embed graphics files
  \DeclareGraphicsExtensions{.eps}     % for pure latex we expect eps files
\fi%

\graphicspath{{figures/}{pictures/}{images/}{./}} % where to search for the images

\usepackage{microtype}                 % use micro-typography (slightly more compact, better to read)
\PassOptionsToPackage{warn}{textcomp}  % to address font issues with \textrightarrow
\usepackage{textcomp}                  % use better special symbols
\usepackage{mathptmx}                  % use matching math font
\usepackage{times}                     % we use Times as the main font
\usepackage{cite}                      % needed to automatically sort the references
\usepackage{tabu}                      % only used for the table example
\usepackage{booktabs}                  % only used for the table example
\onlineid{0}

\usepackage[table]{xcolor}
\usepackage{paralist}
\usepackage{enumitem}
\usepackage{soul}
\usepackage{tabularx}
\usepackage{csquotes}

\usepackage{lettrine}
\usepackage{floatflt}
\usepackage{wrapfig}
\newsavebox\curwrapfig
\makeatletter
\long\def\wrapfiguresafe#1#2#3{%
  \sbox\curwrapfig{#3}%
  \par\penalty-100%
  \begingroup % preserve \dimen@
    \dimen@\pagegoal \advance\dimen@-\pagetotal % space left
    \advance\dimen@-\baselineskip % allow an extra line
    \ifdim \ht\curwrapfig>\dimen@ % not enough space left
      \break%
    \fi%
  \endgroup%
  \begin{wrapfigure}{#1}{#2}%
    \usebox\curwrapfig%
  \end{wrapfigure}%
}
\makeatother
\usepackage{etoolbox}%to help things be more robust /protect'ion - allowing newrobustcmd to be used

\usepackage{float}% so we can use [H]

\definecolor{buff}{RGB}{238,232,170}
\definecolor{myGray}{RGB}{135,206,250}
\definecolor{myRed}{RGB}{165,31,39}
\colorlet{myOrange}{orange!55}
\colorlet{myOrangeFrame}{orange!50!black}

\definecolor{evfpurple}{RGB}{204,153,255}
\definecolor{evfgrey}{RGB}{225,225,225}
\definecolor{evfyellow}{RGB}{255,230,153}
\definecolor{evfgreen}{RGB}{197,224,180}

\definecolor{evforange}{RGB}{226, 137, 27}

\definecolor{evfdarkblue}{RGB}{52, 101, 181}

\definecolor{ABC}{RGB}{255,0,0}
\definecolor{JCR}{RGB}{0,0,205}%{0,0,205}
\definecolor{evfpurple}{RGB}{205,100,205}

\def\mycmd{1}
\if\mycmd1
    \usepackage[switch]{lineno}
  \else
    
\fi
\newrobustcmd{\phaseONEitem}{{\raisebox{\dimexpr-.6\height+.3\baselineskip}{{\includegraphics[width=3mm]{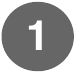}}}}}
\newrobustcmd{\phaseTWOitem}{{\raisebox{\dimexpr-.6\height+.3\baselineskip}{{\includegraphics[width=3mm]{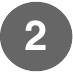}}}}}
\newrobustcmd{\phaseTHREEitem}{{\raisebox{\dimexpr-.6\height+.3\baselineskip}{{\includegraphics[width=3mm]{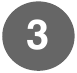}}}}}

\PassOptionsToPackage{unicode}{hyperref}
\PassOptionsToPackage{naturalnames}{hyperref}

\vgtccategory{application/design study}
\vgtcpapertype{design study}

\title{Explanatory Journeys: Visualising to Understand and Explain \\ Administrative Justice Paths of Redress}%
\author{Jonathan C. Roberts, \textit{Member, IEEE}, 
Peter Butcher, Ann Sherlock, Sarah Nason}
\authorfooter{
\item
 Roberts, Butcher, Sherlock, Nason are with Bangor University. E-mail:{j.c.roberts, p.butcher,ann.sherlock,s.nason}@bangor.ac.uk.
 }

\shortauthortitle{Explanatory journeys}
\abstract{Administrative justice concerns the relationships between individuals and the state. It includes redress and complaints on decisions of a child’s education, social care, licensing, planning, environment, housing and homelessness. 
However, if someone has a complaint or an issue, it is challenging for people to understand different possible redress paths and explore what path is suitable for their situation. Explanatory visualisation has the potential to display these paths of redress in a clear way, such that people can see, understand and explore their options. The visualisation challenge is further complicated because information is spread across many documents, laws, guidance and policies and requires judicial interpretation. Consequently, there is not a single database of paths of redress. In this work we present how we have co-designed a system to visualise administrative justice paths of redress. Simultaneously, we classify, collate and organise the underpinning data, from expert workshops, heuristic evaluation and expert critical reflection. We make four contributions: (i) an application design study of the explanatory visualisation tool (Artemus), (ii) coordinated and co-design approach to aggregating the data, (iii) two in-depth case studies in housing and education demonstrating explanatory paths of redress in administrative law, and (iv) reflections on the expert co-design process and expert data gathering and explanatory visualisation for administrative justice and law.} % end of abstract


\teaser{
  \centering
  \includegraphics[width=\linewidth]{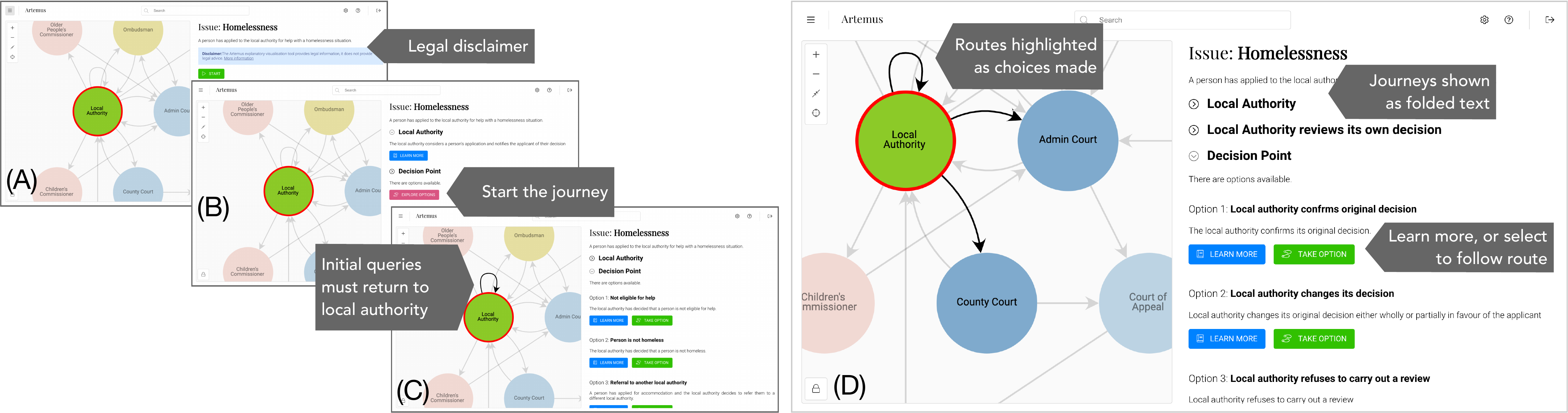}
  \caption{Through co-design, workshops with lawyers, judges, charities, Ombudsmen and so on, we ascertain redress routes, write example scenarios and develop Artemus: our administrative justice explanatory visualisation tool. Focusing on housing and homelessness and education, people learn about possible redress options, e.g., someone who has been made homeless (A), must first go to the local authority (B) and (C), if this redress fails they could go to county or administrative courts (D).}
  \label{fig:teaser}
}

\vgtcinsertpkg

\hypersetup{
    colorlinks=true,
    linkcolor=black,
    citecolor=black,
    filecolor=black,
    urlcolor=black,
}

\begin{document}


\firstsection{Introduction}

\maketitle
Understanding administrative law is important for the general public, especially if they want a remedy or compensation for a decision or grievance. But navigating different possibilities and options is a huge challenge. Even for lawyers it can be tough, because of the intricacies of how laws are created. 
Administrative justice, sometimes known as public law, concerns the relationships between individuals and the state. Public law governs how public bodies act, to ensure that these bodies behave in a legal and fair way. The importance of administrative justice cannot be underestimated, everyday throughout the world, humans interact with the state; about decisions of a child's education, relative's social care, licensing, planning, environment, housing and homelessness. But the public are often unaware of basic legal principles, much less understand how to make a complaint on decisions that affect their lives. With different options for redress, it can be difficult to understand which of the multiple routes they should take. Taking one route may inhibit them taking another; one form of redress may need to be tried and rejected before certain options become possible, or they may decide to take no action. 

Explanatory visualisation techniques have potential to present clear information while helping people understand different choices of redress. Initially we focused on policy makers and legislators, but though consultation our emphasis shifted to explaining redress paths to an aggrieved person sitting alone or with an advisor. Well crafted diagrams, maps and charts can make the information clearer to interpret, easier to understand, and more succinctly demonstrate the concepts~\cite{rosman2013visualizing}. But how do we visualise this information? How do we show different routes of redress? How can someone see every option, navigate through the routes and understand relevant information to their situation? In this article we present how we designed and built a visualisation tool (Artemus) to display different pathways, and help people navigate options for redress. 

\begin{figure*}[t]
    \centering
    \includegraphics[width=\textwidth]{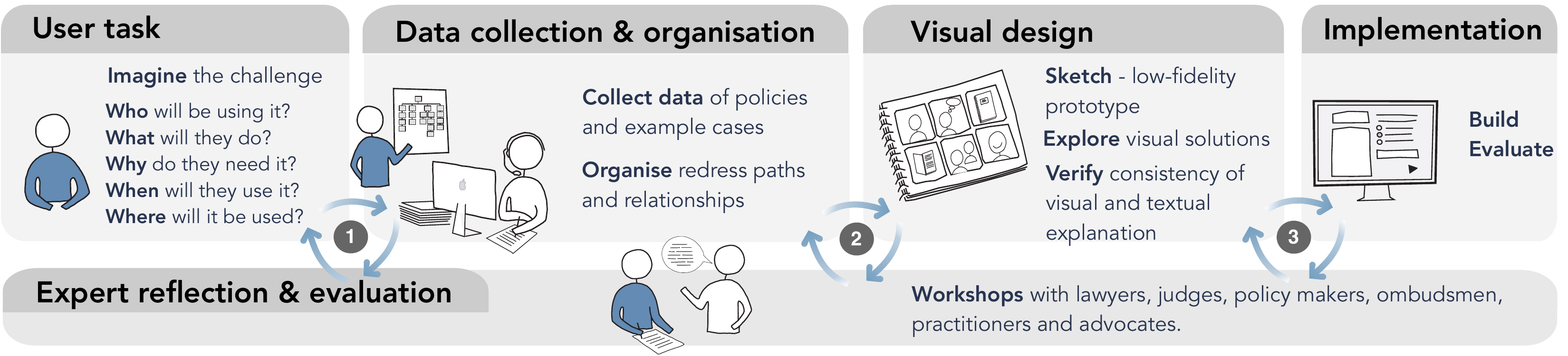}
    \caption{To develop the explanatory visualisation we consider five core concepts. The task for the user; data collection and its organisation; visual design of the solution; implementation. All these processes are underpinned by expert reflection and evaluation. Using an iterative approach, we explain three cycles of design. Phase \texorpdfstring{\phaseONEitem}{1} focuses on the task, data collection and reflection. Phase \texorpdfstring{\phaseTWOitem}{2} confirms the collected data and its classification. Phase \texorpdfstring{\phaseTHREEitem}{3} focuses on the design, implementation and evaluation of the tool.}
    \label{fig:FlowDiagram}
\end{figure*}

In addition to the challenge of visualising paths, there is the question of gathering and processing data. It is not a matter of plugging in a visualisation tool to a database, as there is no single database of information. Legal principles are complex, information held across laws, and discernment is needed to make distinction between law and policy, rules and discretionary decision-making. 
Furthermore the system of institutions and remedies has grown up piecemeal in a very ad hoc manner and presents the individual with a plethora of different bodies, processes and laws. Navigating the different possibilities can be a huge challenge. 
Consequently, the process of developing explanatory systems must be synchronised with the process of collating, classifying and simplifying data. The process involves assembling information from various sources, including documents and expert opinions, classifying and sorting the data into something that can be stored and used in a visualisation process. Of course, because people will rely on these systems to make decisions, they need to be correct, unbiased and exemplify real situations; hence the data is judged by expert critical reflection that we obtained through four expert workshops and direct interaction with other stakeholders. Visual methods, and visualisation design strategies, including sketching, expert workshops, and heuristic evaluation can help to collate and organise the required data. 

We focus on Welsh law and in particular on practices in housing and education. This focus provides a convenient test bed for our data gathering and visualisation demonstrator. In the UK, administrative law is a mix of common laws of England and Wales, devolved and national law. From previous work we had links across the UK with judges, legislators, practitioners and administrative justice educators~\cite{nason2016reconstructing,nason2013regionalisation}. Consequently we were able to hold workshops with relevant and committed stakeholders. We performed a comprehensive review of public administration in Wales, developed the pathways, conducted several workshops to collaboratively review and adapt the administrative justice paths of redress, and developed associated texts as case studies to exemplify these examples. There is a strong need to be clear and consistent in both languages. We use bilingual drafting, to co-draft what is written in English and Welsh at the same time, rather than translating from one to the other.
We make four contributions:
\begin{enumerate}[left=0pt,nosep]
\item An application design study of an explanatory visualisation tool (Artemus). 
\item Coordinated and co-design approach to developing the visualisation coincidentally with the data paths and exemplar case studies. 
\item Administrative justice case studies in housing and education to explain administrative justice paths of redress. 
\item Reflections and guidelines on developing explanatory visualisation applications. 
\end{enumerate}

\section{Study approach}
The administrative justice system ensures that public decision-making is lawful and fair. It provides systems to ensure that duties and powers of different public bodies are carried out properly. These systems help people address issues if they believe they have been treated unfairly or unlawfully. Our research set out to examine administrative justice in Wales and make recommendations for improving it.
By capturing experiences and using visualisation approaches, we examined the Welsh approach to administrative justice and its capacity to respond to change. To recommend improvements of the administrative justice system, we needed to perform extensive research into general administrative justice and administrative law. This data needed to be organised and tested with expert reflection. It was necessary that the relationships that we were mapping were correct. We also wanted to create examples, and scenarios, which were taken from real-life situations, consequently we needed to collate examples from practitioners.

Our study approach was agile and iterative. We used a collaborative, co-design strategy, whereby we researched, designed and implemented at the same time. We followed the design cycles from Roberts et al.\ \cite{RobertsHeadleandRitsos16_FDS_TVCG,RobertsHeadleandRitsos2017_FDS_BOOK} ever-improving and cycling from low-fidelity designs that are evaluated by practitioners, to developing better and more complete solutions which are likewise evaluated. During our project we moved seamlessly between data-collection and organisation of the data, to expert reflection, visual design and implementation. However, on reflection we realised that there were three main phases. We show these parts diagrammatically in \autoref{fig:FlowDiagram}, along with the three phases, and use this structure to organise the paper.
\begin{itemize}[nosep,topsep=1mm,left=4pt]
    \item[\phaseONEitem]  In \textbf{phase~1}, we considered the user, comprehensively investigated  administrative justice and started to develop initial networks. Before the project started in earnest we performed a stakeholder survey to evaluate the need (achieving 60\% replies). 100\% agreed that processes for redress in Wales needed rationalisation, 57\% indicated that the current arrangements for allocating grievances were not clear, demonstrating there was utility for the project.
    \item[\phaseTWOitem] \textbf{Phase~2} focused on validating and improving our outline networks, which we achieved through a series of in-person workshops and direct engagement with experts and practitioners. We produced low-fidelity network diagrams in Microsoft Word to demonstrate the networks and specific pathways of redress, which were used in the workshops. 
    \item[\phaseTHREEitem] In \textbf{phase~3} we performed a deep-dive into alternative design ideas, sketched alternative designs and implemented the Artemus explanatory visualisation tool; testing and validating it with experts.
\end{itemize}

\section{Background and related work}
\label{SEC:BackgroundAndRelatedWork}
The purpose of developing an \textit{explanatory visualisation} is to explain data and processes. However, explanatory visualisation is only one style of many: developers may create a visualisation to allow users to \textit{explore} data, \textit{present} results, tell stories and show findings, or  \textit{explain} phenomena~\cite{RobertsETAL2018_EVF,brooks1993,KosaraETAL2016}. Considering these three styles, respectively, a developer may say: ``this visualisation allows me to explore data'', ``this visualisation enables you to see what I see'', or ``what do you see and learn from this display?''~\cite{brooks1993}. When a developer designs a visualisation they must understand the purpose of their visual output. They need to understand where it will be used, the skills of the user who will use it, and the purpose of the visual display~\cite{RobertsHeadleandRitsos2017_FDS_BOOK}. All these factors affect decisions, that the developer needs to make, on the design of their visualisation, how the user interacts with the visualisation, appearance, choices of which software is used to implement it, and where and how it is published. The designer will make different decisions when creating an exploratory visualisation in comparison to an explanatory one.

With \textit{exploratory} systems, researchers want to discover relationships and understand data. Tukey's championing of exploratory data analysis~\cite{Tukey1977exploratory} sparked off the development of statistical and exploratory tools and libraries~\cite{stasko2008jigsaw,Weaver2004,bostock2011d3,IhakaGentleman1996_R}, and interactive and exploratory query based visualisation techniques such as multiple linked views~\cite{Roberts2007,AlManeeaRoberts2019Quantifying}, dynamic queries~\cite{ShneidermanWilliamsonAhlberg1992DynamicQueries}, cross filtering~\cite{Weaver2004} and visual comparison~\cite{GleicherETAL_VisualComparison2011}.
Through exploration developers can test hypothesis, discover outliers, trends, or unexpected patterns~\cite{KosaraETAL2016}. 
For example, a researcher modelling sediment transport, would change different parameters of the model, and visualise and compare alternative outcomes~\cite{George_Evn2014}. The quantity of research in this area is huge; academic researchers have investigated, developed solutions for, and published many articles on exploratory visualisation techniques, data analysis and systems that are highly interactive, consequently there are many review papers~\cite{LipsaEtAL2012,Roberts2007,zudilova2009overview,sun2013survey}.

When researchers and practitioners \textit{present} a visualisation they are storytelling~\cite{SegelHeer2010,kosara2013storytelling}. To present an effective story, developers must have a deep understanding and comprehension of their data, and show it in a way that guides the observer through the material. For instance, an academic would learn the material, summarise it and lead the students through the core concepts. Or a journalist would write a story and create a visualisation from data. Often these visualisations are accompanied with different channels of information. Each channel needs to express the ideas homogeneously; titles, text, verbal commentary, visualisations, animations and so on, all need to present a consistent and coherent story.
The area of data-storytelling, narrative visualisation, data journalism, data presentation and so on, is popular in the academic domain~\cite{hullman2013deeper,HullmanDiakopoulos_VisRhetoric2011,liem2020structure2020} but many of the concepts and strategies have been driven by social media influencers, journalists and publishers. 

With \textit{explanatory} visualisation, the goal is to provide both a way to present facts and allow users to explore it so that they can develop a deeper understanding of the information. It has three primary goals: (i) educate, upskill and instruct, (ii) elaborate concepts and processes instead of focusing on data sets, and (iii) to elucidate processes and explain what it is, why it happens and how it relates to other concepts~\cite{RobertsETAL2018_EVF}. The process requires developers to explicate and draw out the meaning of the underpinning data, which is often not clearly defined~\cite{oliveira2015use}.
Far less research has been published on explanatory visualisation~\cite{RobertsETAL2018_EVF}. 
It is clear, however, that explanatory presentation is popular, with many public bodies and industries creating explainer videos, bloggers creating videos to explain how to make something or repair equipment, and interactive explanation in journalism, with the public co-writing or creating interactive visualisations and explaining concepts.
Explanatory visualisation is subtly different to (but can utilise) data-storytelling or narrative visualisation methods. For example, a journalist may create a narrative visualisation to describe how a pandemic spread, using visualisations to quantify the hospitalisations, deaths, recoveries and so on. Explanatory visualisations or other explanatory cues could be used alongside, to explain how data is captured, numbers are calculated or what specific words mean.

Explanatory visualisations have been applied to many different topics, used to explain mathematics principles~\cite{Guzman2002ft,arcavi2003role}, general relativity~\cite{WeiskopfRelativity2006},
communicate geological concepts~\cite{Natali_ETAL2012},
teach algorithms~\cite{RobertsETAL2018_EVF}, present definitions and explain words~\cite{roberts2018visualisation,RobertsETAL_MultViewsMeaningsWords2019}, explain data-structures and other computing concepts~\cite{fouh2012role,sorva2013review,naps2002exploring}, or general scientific principles \cite{Tufte1997,LipsaEtAL2012}. 
Explanations have been used as part of interactive tools to explain different concepts. For instance, in multiple view systems it can be difficult to understand what views are linked~\cite{Roberts2007}, which can be explained with overlaid metaviews~\cite{kubota2014evolve,Weaver-MetaVisualisation2005}. Similarly variable sensitivity information~\cite{spence2004sensitivity} or trails of user activity could be visualised to explain where users could or have navigated, allowing them to return to previous versions~\cite{BavoilETAL_VisTrails2005}.  
Early explanatory seminal work in the late 1990s and early 2000s principally used animation to help explain concepts~\cite{Urquiza2009survey,shaffer2010algorithm}. Similar techniques are used today by teachers, as exemplified by the quantity of explainer videos that use animation and can be viewed from an Internet search of ``visualisation of sorting algorithms''. 

Educators often use visual explanation methods, particularly teaching scientific facts~\cite{RobertsETAL2018_EVF,oliveira2015use}. 
They use visual representations accompanied with commentary or other discourse to explain core concepts. Many authors encourage a multimodal approach~\cite{adadan2013using,ainsworth2006deft,kozma2003material}, where pictures are used alongside verbal description, or several pictures are used to show alternative viewpoints. Whatever modalities used, many educators include visual pictures to illustrate their points. For instance, explaining how the heart works to medical students is much easier with a visual diagram~\cite{butcher2006learning}, or explaining inter-molecular interactions are easier with videos~\cite{oliveira2015use}, and using visualisations to explain chemistry constructs is ``central to the development of chemical understanding''~\cite{kozma2005students_ChemistryVis}.

While educators use models, pictures, animations and so on to explain complicated processes and models, their explanatory depictions need to be more than merely `explaining the results'~\cite{BraatenWindschitl_2011_SciExplanation}. If the concepts are easy to learn, teachers will explain them orally. Consequently, learners need to concentrate and relate the new information they hear to knowledge they already know~\cite{biggs1999student}. They are juggling, in their mind, the new information, placing it alongside information they already know, and trying to resolve any contradictions or questions they have about the new knowledge. Not everyone has the same background, or develops understanding in the same way, consequently it is important to explain in different ways and use several strategies. Treagust and Harrison~\cite{TreagustHarrison2000_ExplanatoryFrameworks} explain that learners create ``explanatory frameworks''. Learners need to see the information in different ways, relate it to knowledge they know, and test their understanding. If the user can see the world from another person's viewpoint, and empathise with their point-of-view, they are more likely to deeply understand the issues involved, and understand how to proceed. Consequently, in our designs we wanted to include alternative views, real-life examples to demonstrate how the knowledge and paths of redress could be applied to the viewers' life, and allow users to interact and explore the information in a guided way.

Explanatory visualisation fits well with our challenge of explaining redress paths in law. We do not want to give users an open-ended exploratory system.  Instead, we need a way to present clear structures, while allowing users to explore and discover relevant information to their personal circumstance to develop a robust understanding of paths of redress. On the other hand, we do not want to merely `present' the information, instead we want them to develop effective `explanatory frameworks' and develop a deep understanding of the material in relation to their situation. Ours is a constructivist approach. We want the public users of our law visualisation tool to have a hands-on experience; permit the public to learn about the different paths of redress, and apply them to their own situation. After interacting with our explanatory visualisation tool, we want the users to have a good understanding of their options, and to have more confidence in their own decision making. It may be that after using our tool people decide to do nothing -- they decide that the right decision has been made in their case. Or they will decide they do need to take action, and be better informed how to move forward and who to contact to help resolve their issue. There would be many ways someone could move forward from this point. In fact new tools are being invented to help people write letters and make complaints, such as through the resolver service (\href{http://resolver.co.uk/}{resolver.co.uk}). However our focus is to explain and inform paths of redress. The person may seek council from the body they have the complaint with, take court action, complain to an ombudsman, take legal action, or receive advice from an advocacy charity~\cite{nason2016reconstructing}. Indeed, they may be using our explanatory tool with an adviser or advocacy service, who can use it to help explain their different options. 

Our focus is to visualise and explain paths of redress, and while there has been less research in this area~\cite{rosman2013visualizing} visualisation techniques have been successfully applied throughout law. 
For instance, visualisations have been created for use in a courtroom to recreate scenes, and annotations of photographs have been used to express where people went~\cite{noond2002visualising}. 
Visualisations have been used to help investigators understand criminal patterns~\cite{BRUNSDON200752_VisCrimePatterns}, showing timelines of activities~\cite{noond2002visualising} or
alternative opinions~\cite{bourne2005visualising}. Where explanatory visualisation has been used, it has been dominated by flowcharts~\cite{mclachlan2020visualisation}, which have been used to visualise argumentation~\cite{oliveira2015use,kirschner2012visualizing} or explain legal instructions to civil servants~\cite{passera2018flowcharts}, network diagrams to visualise policy making~\cite{Burkhardt2015ExplorativeVO,KohlhammerVisualisingPolicyModeling2012} and  schematic diagrams to explain processes in contracts~\cite{Passera2012_ContractUsabilityWithVisualisation}.
This prior work led to the network metaphor being included in the design process.

\section{Phase \texorpdfstring{\phaseONEitem}{1} -- User task, data collection, reflection}
\begin{figure}[H]
    \centering
    \includegraphics[width=\columnwidth]{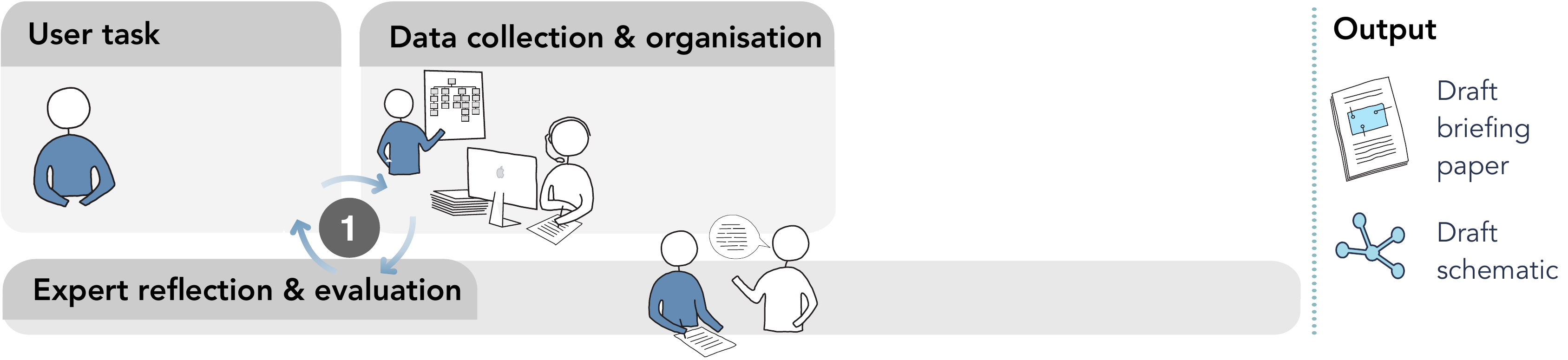}
    \vspace{-5mm}
    \caption{In Phase~1 we reflect on the task, actors of the system, collected and organised information on housing, homelessness and education. We delivered a briefing paper and draft schematic.}
    \label{fig:FlowPhase1}
\end{figure}

In the first phase we  collected data, started to classify the paths between different institutions and organisations, designed a first network diagram, and reflected on the user task. Our goal, at this stage, was to create draft documents and information that others could critique, to guarantee that the information we are explaining is correct. We drew upon our expert knowledge, spoke with practitioners and other experts. We output a briefing paper, with accompanying schematic diagram, as shown in \autoref{fig:FlowPhase1}. In this section we start by describing the data and then go onto the user task. 

\subsection{Data collection and organisation}
We used two specific case-study areas to enable us to support our general conclusions and recommendations with concrete examples. 
The areas selected were \textbf{social housing and homelessness}, and \textbf{primary and secondary maintained (public) education}. These are both areas where power to make the laws has been devolved to the Welsh Parliament (previously the National Assembly for Wales) and therefore any recommendations for change made by us could be taken up by the Welsh Government. They are both areas where there is a plethora of general and specialist bodies with a role in dispute resolution. For example, in education, there are governing bodies, specialist exclusion and admission appeal panels, and an Education Tribunal. This is in addition to the general court system, and the `integrity' bodies such as the Public Services Ombudsman for Wales, the Children’s Commissioner for Wales, the Welsh Language Commissioner and the Future Generations Commissioner for Wales. Looking into the case study areas in detail provided an opportunity for us to test and support our general views about the administrative justice system at large. 

But there are huge complexities in law that we needed to understand. Successive statutes on devolution have given the Welsh Parliament the power to make its own legislation on specified topics, known first as \textit{measures} and now as \textit{acts}. Housing and education are topics on which the Welsh Parliament has the power to legislate. However, pre-devolution legislation made by the UK Parliament is still applicable in some instances, as are some of the regulations made by UK Government ministers under the authority of those UK Acts of Parliament. Frequently, an \textit{act} is amended on many occasions so there is a challenge in knowing whether a provision is fully up to date. Accordingly, the applicable law comes from a range of legal instruments: acts of the UK Parliament to the extent to which they still apply in Wales, \textit{measures} and \textit{acts} of the Welsh Parliament, regulations made by UK Government ministers to the extent to which they still apply in Wales and regulations made by the Welsh ministers under the authority of UK acts or Welsh measures or acts. The law is frequently accompanied by statutory guidance issued by ministers where the requirement is that the relevant bodies have to `have regard' to what it says: this means that they are not strictly obliged to follow it, but they need to show that they considered what it says and have a good reason if they decide not to follow it. In addition to the specific law on housing and education, it is also necessary to be aware of the general law applying to remedies and of cross-cutting law on issues such as human rights and anti-discrimination law. Overall, this amounts to a complex web of different instruments from different sources and with different legal status, not easily understood or accessible by many lay people. 

In relation to housing and education, we examined legislation, the existing limited commentary on that law in Wales, case law, previous research, statistics and reports from government, representative organisations, charities, relevant tribunals, local authorities, and the work of institutions such as the Children’s Commissioner for Wales, the Older People's Commissioner for Wales, The Welsh Language Commissioner, the Future Generations Commissioner for Wales and the Public Services Ombudsman for Wales. We compiled a register of institutions and processes relevant to challenging initial public decision-making in these areas, and the more general laws on dispute resolution, cross-cutting issues such as human rights and anti-discrimination law. This enabled us to compile a briefing paper for each of the two areas (housing/homelessness and education) with an account of the relevant law, areas of weakness or contention, possible disputes that could arise for an individual, institutions that could be involved in providing redress, and different redress outcomes that could be pursued by an individual. 

The briefing paper contained many examples of redress. For illustration we can consider the situation that someone was made homeless, and they perceive that a local authority has breached a duty owed to them to prevent them becoming homeless, or a duty to accommodate them, then they would have several options: 
\newenvironment{redress}[1]% environment name
{% begin code
  \par\vspace{.1\baselineskip}\noindent
  \hfill\begin{minipage}{\dimexpr\columnwidth}
  \textbf{\itshape #1}\itshape~``\noindent\ignorespaces
}%
{% end code
\ignorespacesafterend'' \par\vspace{.10\baselineskip}\noindent\normalfont\ignorespacesafterend\end{minipage}
}
\begin{redress}{Redress one (reconsideration by the Local Authority):}
You should first ask the local authority to reconsider its decision. There are a small number of types of decision about homelessness where you do not have the right to ask for a reconsideration.%
\end{redress}
\begin{redress}{Redress two (appeal to the County Court):} 
After you have asked the local authority to reconsider its decision, if you still feel it has made the wrong decision you can appeal to the County Court.%
\end{redress}
\begin{redress}{Redress three (Court of appeal):}
If you think the County Court has made the wrong decision you may be able to appeal to the England and Wales Court of Appeal.%
\end{redress}
\begin{redress}{Redress four (judicial review):}
If you think the County Court reached its decision in a way that was procedurally unfair, for example you feel the judge did not properly take into account your evidence, you may be able to seek judicial review in the Administrative Court. This review will be about the way the County Court decision was reached -- not about the decision itself. (This route is rarely used).%
\end{redress}
\begin{redress}{Redress Ombudsman:}
You can complain to the Public Services Ombudsman for Wales about something called `maladministration'. This is where you feel you have been treated unfairly or received a bad service from the local authority. Sometimes this kind of unfairness may also mean that the local authority has broken the law. If you think the local authority has broken the law, and you have a right to appeal or seek judicial review in a court, then usually the Ombudsman will not be able to look into your complaint at the same time.%
\end{redress}
\par\vspace{.2\baselineskip}\noindent

In the briefing paper we included more examples and paths. It is important to note that not all the paths would be relevant, and choosing to take one option may inhibit others. In fact, during this phase we highlighted several judicial challenges with the current system, that we highlight in the final report~\cite{nason2020public}. 
For instance, with the `Redress Ombudsman' example, academics, lawyers, judges and others struggle to make a clear distinction between matters for the Ombudsman and matters for the courts. Consequently we have suggested a reform in Wales, since supported by the ombudsman and the Commission on Justice in Wales, for more flexibility that allows the ombudsman to refer a point of law to the courts and the courts to engage the ombudsman. This leads to more flexibility in redress, but not in a way that adds further complexity to pathways for individuals trying to navigate them.

\subsection{Data organisation, diagramming and reflection}
The networks  we are investigating (pathways of redress) are not simple structures. They contain uncertain connections, alternative routes, multivariate variables and uncertainty. Our goal was to explore major connections, investigate alternative layout strategies. To create a system that can help people understand these structures, learn about possibilities and fit the knowledge with information, or correct what they already know. The visual structures can also help identify problems with legislative coherence that otherwise would not be recognised.

\begin{figure}[h]
    \centering
    \includegraphics[width=0.9\columnwidth]{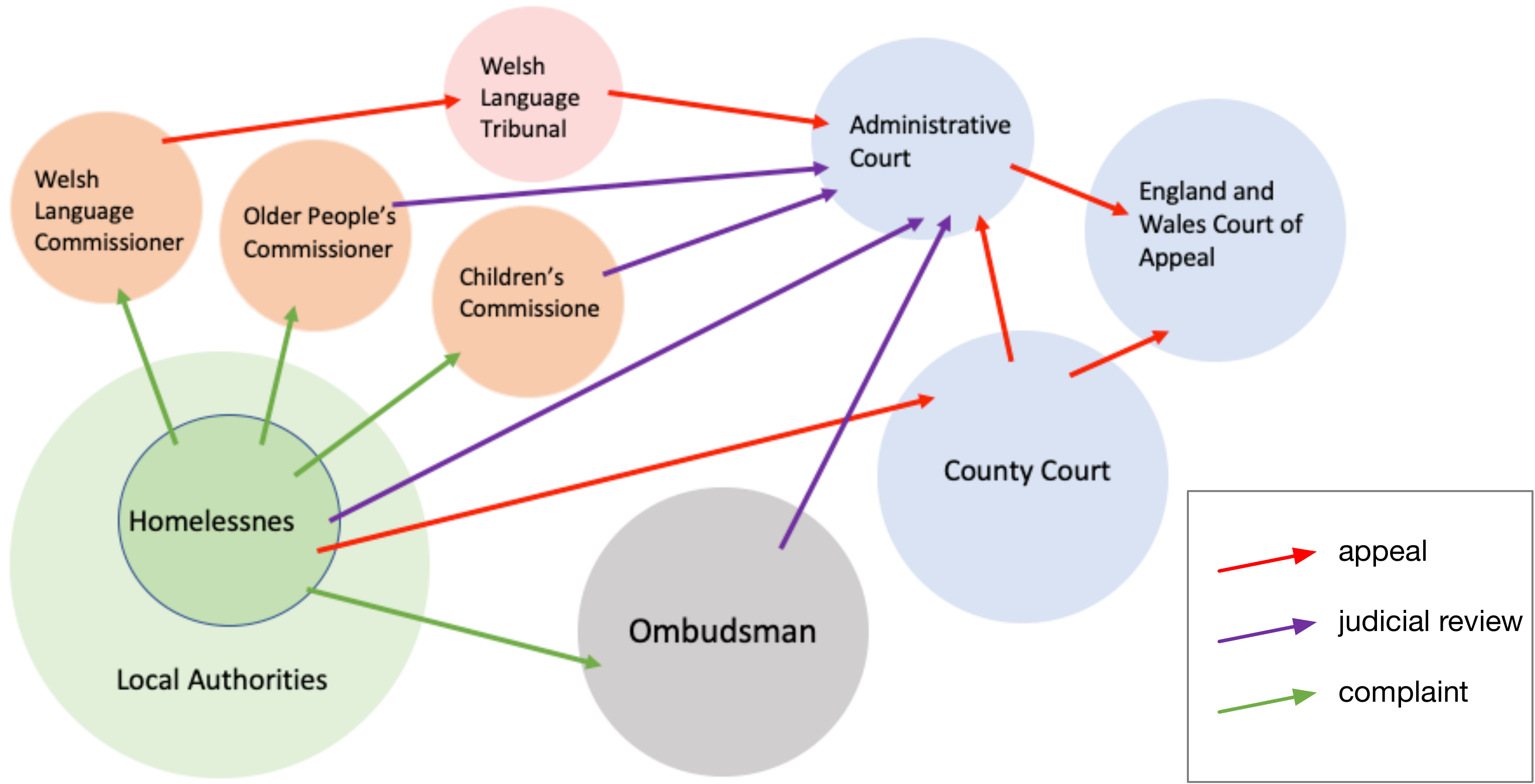}
    \caption{Early explanatory schematic diagram (of the housing/home\-less\-ness case study). The image was developed in phase~1, shared in Microsoft Word, and used as part of the critical engagement with stakeholders.\vspace{-3mm}}
    \label{fig:HomelessDraft}
\end{figure}

In addition to the briefing paper we started to map these routes as a hand-crafted network diagram (\autoref{fig:HomelessDraft}). 
Throughout the project, our approach was one of going back and forwards between the legal data and the practical experiences of how people use and experience the legal system. For the mapping, this involved identifying the points at which an individual would become engaged with a problem in the administrative justice system and enter one of our interactive maps.
We followed a process of working from examples and building the network of connections from the laws and procedures. It was for the convenience of sharing with our experts that we chose to draft these routes in Microsoft Word diagrams. This meant we could include the network diagrams in Word documents and PowerPoint presentations and share them with our experts, who could add comments and amend as appropriate. These diagrams were `working diagrams'; not final visualisations. In fact, we wanted people to discuss the routes and confirm or deny the pathways. 

At this stage we also started to add extra subtleties to the diagrams, such as colours to represent types of redress: 
green for a complaint to the ombudsman, red to represent an appeal to a court or tribunal, and judicial review in purple. Referencing the diagram in~\autoref{fig:HomelessDraft}, a typical reflection, by the main law experts in the project and from expert feedback, went as follows:
\begin{redress}{Local Authority homelessness duties.}
Consider a person who is concerned about a specific decision that the local authority has made (e.g., not to accommodate them) or are they dissatisfied with the service they’ve received (e.g., someone has been rude on the phone, or sent them the wrong information etc), or is it both these things.%
\end{redress}
\begin{redress}{Decision for the complainant to make.}
This is where you have to decide are you going down a purple/red route or a green route, or both. It’s the both part that can get complicated, I think the only way we could really deal with this is try to give a specific example of when someone can do both -- through a link or `pop up' thing 
so people can see the rare circumstances where they might be able to do both.% 
\end{redress}
\begin{redress}{Options.} 
There are different routes to redress depending on whether a person wishes to challenge a specific decision that a local authority has made in connection with duties that they believe the authority owes to them, or whether the person feels they have been treated unfairly (but not unlawfully) or received a poor service.% 
\end{redress}
\begin{redress}{Decisions to be made.} 
In this example we can split specific decisions into Type A (one redress pathway) and Type B (a different redress pathway). You normally can't have a red and a purple route -- it's either one or the other. Historically purple has always been available as a matter of common law (judge made law) in some areas Westminster Parliament or National Assembly have created specific statutory appeal (red) routes -- but the purple route remains available for all areas where a red route hasn't been specifically provided for by a statute.%
\end{redress}\par\vspace{.1\baselineskip}\noindent

It was imperative to have these dialogues between the experts. We were identifying potential pathways, evidencing them with real-life examples, gathering the draft text for the final explanatory visualisation tool, and ascertaining potential design ideas.

\subsection{User task -- who, what, why, when, where}
\label{sec:UserTask}
Understanding the audience is one of the most important challenges to explanatory visualisation. If we, as developers, do not understand who will be using it, their skills and where they are at, there is little chance that we will help them learn. They will not use the tool, because they will be confused. To understand the user requirements we chose to focus on the five w's method. We chose this methodology because it is easily achieved, well known by the collaborators, and we have used it successfully before~\cite{RobertsHeadleandRitsos2017_FDS_BOOK}. Our discussions over the `user' were held in tandem with the development of the networks and the examples.

\textbf{Who would use the explanatory visualisation?} To understand who is involved, we looked at the different stakeholders who have interest in this type of visualisation. Choosing what information is to be explained and who would be the principal benefactor of the visualisation is important. These decisions would alter the type of explanatory visualisation that we create. We created several outputs (reports, briefs, presentations) for different groups, to tailor the information, the method and language used. For instance, explaining the processes to a professional law maker, required different language in comparison to explaining the processes to a lay person.
Our initial focus was to create the explanatory visualisation for the law makers and policy creators, however after discussion with stakeholders and sitting in an emotionally charged meeting with people who believe they had been aggrieved, we chose to focus the explanatory visualisation tool for the `aggrieved person', or an advisor who would be sitting alongside the aggrieved individual. With substantial cuts to legal aid, few people are eligible for funded legal aid advice. However, both digital competence/digital literacy, among the population (of England and Wales), and broadband connectivity have increased. People are seeking legal-related help who are more literate generally, more digitally literate, and have a better understanding of their rights; but still need advice and support from a specialist to act on the information available to them. Vulnerable groups need to be properly supported by advisers and the visual approach of Artemus may well be easier for them to grasp than reams of text~\cite{LundgardLeeArvind2019}. Subsequently, we refined specific outputs for specific purposes: the written reports to the policy makers, and the explanatory visualisation to the general public.

\textbf{What to explain?} To understand what to visualise and explain we needed to analyse the data in detail. In addition to the bodies, organisations, routes of redress, and the case study examples, we estimated quantities of each redress type in Wales, and success rates and frequency of path use. For instance, we know that Redress three (Court of appeal) is rare with only a handful of appeals of this type occurring. While it could be possible to display this uncertain information, through expert consultation, we decided that this information could introduce biases in how people perceive the processes. Subsequently, we decided to not visualise additional metrics, but to focus on the pathways (nodes and edges of the network) and real-life examples. 
We considered long and hard what the \textit{nodes} and \textit{edges} meant, and how they would be used to explain the information. One solution could be to display the links in the network to represent how one body interacts or reviews another. For example, the ombudsman not only looks into complaints about public services but also investigates whether government bodies have violated their own code of conduct. Potentially we could explain other relationships, such as ``the body belongs to'', or ``the body advises'' and so on, which would create hierarchical structures. However redress is an emotive subject and concluded that it would be better for the individual to relate to the information personally. We decided that the routes of redress are a personal tour. The network is viewed from a personal standpoint: \textit{nodes} as bodies or organisations; \textit{edges} give direction and mean that the person would ``take an issue to the next body''. 

\textbf{Why would people use it?} It is obvious that the goal is to educate, elaborate and elucidate, but it is important to delve into the situation of why they would use it. We are dealing with decisions over peoples' lives and complaints thereof. Redress is personal and people respond emotionally. Consequently the information needs to be clear so it can be used in their time of need. It should help them positively, but also be something that reassures them that they can move forward from this situation. Educate and inform them in a unbiased way. 

\textbf{When and where would the explanatory visualisation be used?} Our vision is that people would investigate the possibilities in their own time, or during a visit to an advisor. Therefore we wanted the tool to be load anywhere and be cross-platform.

\section{Phase \texorpdfstring{\phaseTWOitem}{2} -- Data collection, design, reflection}
\begin{figure}[h!]
    \centering
    \includegraphics[width=\columnwidth]{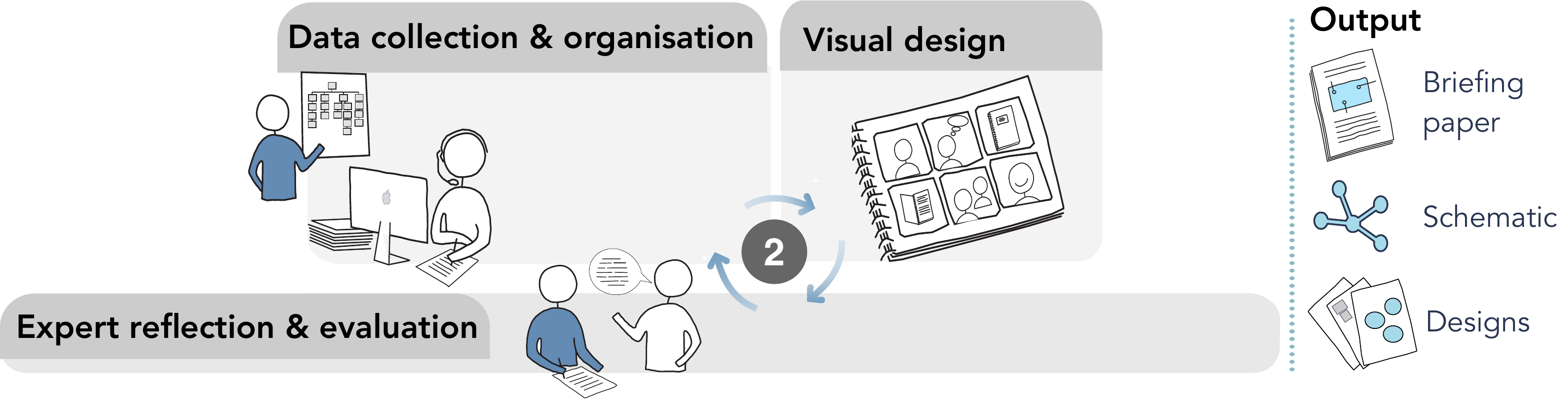}
    \vspace{-5mm}
    \caption{In Phase~2 we focus on (i) adding detail and to the housing/homeless and education examples, (ii) validated the scenarios and expanded the schematic diagram with expert feedback at the workshops, and (iii) considered different design solutions for the final explanatory visualisation.}\vspace{-2mm}
    \label{fig:FlowPhase2}
\end{figure}

The aim of the second phase was to (i) develop and confirm that the examples and the pathways are correct, (ii) draft the visualisation designs and discuss them with stakeholders, and (iii) increase detail on the schematic diagram. We held two workshops for each domain: housing/homelessness and education. We explain the first set of workshops in this section (\autoref{fig:FlowPhase2}), and the second in Phase~3 (\autoref{sec:phase3}, \autoref{fig:FlowPhase3}).

\subsection{Expert reflection and evaluation (the first workshops)}
\label{sec:FirstWorkshop}
The workshops involved over 100 people from different sectors, were aimed at professionals, and we used bilingual (Welsh, English) adverts, handouts and PowerPoint slides. The workshops included specialist lawyers, government officials, charities, representative organisations, local authorities, a range of advocacy and advice providers, the Law Commission (a specialist law reform body for England and Wales), judges, students and representatives from the Welsh Tribunals, from the Public Services Ombudsman for Wales and Welsh Commissioners. We invited people to attend workshops largely based on our existing networks, including connections through our membership of the UK’s Administrative Justice Institute and Administrative Justice Council. In addition, workshops were advertised through the Welsh Government’s National Advice Network (a 150-member network of advice and advocacy providers across Wales), through the organisation Public Law Wales (a representative organisation for public law practitioners in Wales), and through the Welsh Local Government Association. We also had specialist participants from each field such as local authority staff, housing association staff, and bodies representing school governors and head teachers. Before the workshops we sent a copy of the basic schematic to the registered participants (\autoref{fig:HomelessDraft} shows the homeless draft network diagram).

We split the workshops into two parts. First we received presentations from professionals and discussed the key administrative justice issues affecting each sector; from legislation, to avoiding disputes, early resolution and different formal methods of dispute resolution, as well as what gives rise to disputes and how to learn from them, and what reforms could be proposed to the sector as a whole. Second, we discussed the pathways as represented on the schematic diagrams and design ideas (\autoref{fig:HomelessDraft}). We shared printed copies of the schematic pathways diagrams, and pinned a large version as a poster on the wall, which participants edited and annotated.

\begin{figure}[h]
    \centering
    \includegraphics[width=0.9\columnwidth]{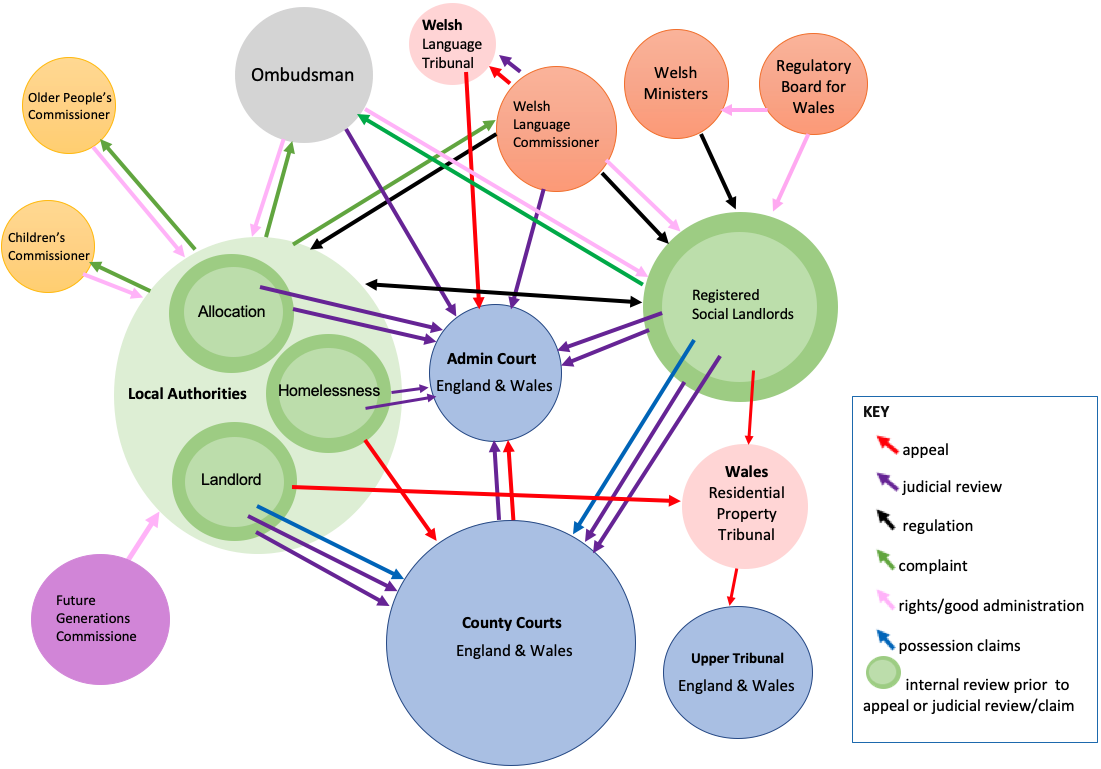}
    \caption{The second draft schematic diagram of the housing/homelessness case study, improving \autoref{fig:HomelessDraft}. It is a functional, work-in-progress, diagram that was developed in phase~2 using Word diagramming, as a convenient way to engage with stakeholders, and make pragmatic decisions of redress paths.  It demonstrates the complexities and thus confirms the need to explain this information an accessible way.}
    \vspace{-2mm}\label{fig:HomelessDraftImproved}
\end{figure}

Feedback at the workshops informed our development of the schematic diagram and allowed us to confirm and adapt the different redress paths. We continued our engagement with stakeholders and conducted specific activities in each sector such as observing legal proceedings, holding specialist focus groups with sector professionals and with parents/carers of children with additional learning needs, and continued to take updates on issues affecting the field of law, policy and practice, alongside Freedom of Information Act requests and surveys. By the start of phase~3 we had created a more detailed version, as shown in \autoref{fig:HomelessDraftImproved}.
In addition to discussing issues and confirming paths, the  workshops highlighted several aspects
% of the redress paths 
that we should explain.
\begin{enumerate}[nosep,topsep=1mm,left=0pt]
    \item `Doing nothing' was an appropriate option, and the explanatory visualisation should make it clear that this is a valid outcome.
    \item Some options may be more difficult to understand and need further explanation. This could be to explain technical language, or to justify or reason that one option is preferred to another, or that taking one route closes down another. 
    \item Some people know more about redress paths open to them, so it would be good to display different levels of abstraction in the examples. The suggestion was to have a `find out more' button, to include detail, reasoning or links to external bodies and advice. 
    \item Each body involved has a different remit and varying levels of cost, formality and accessibility, and the outcomes may differ. It is important to emphasise differences between bodies and their requirements. For example, a court or tribunal will deliver a binding ruling whereas the ombudsman will deliver a recommendation only, albeit one that is almost always followed.  Where there are choices, an individual can explore different options and institutions and the map helps them to appreciate each possible outcome.
\end{enumerate} 

\noindent Through our workshops it became clear that if our design was to focus on the user, as a member of the general public or even as a generalist advice provider, we needed to make clear we were giving `legal information' and not providing `legal advice' (legal advice providers must be properly regulated and insured). One way to be clear that information and not advice was our goal, was not to ask users for personal information and to base our pathways on commonly occurring `use cases' (the most common pathways travelled through the administrative justice system). The mapping tool would still be bespoke to the user's decisions/choices at various stages, but it would not collect any personal data. We also researched the range of disclaimers used in other legal information and legal educational websites, and from workshop feedback determined to provide a disclaimer on the `landing page' of Artemus itself, which is then repeated within pathways when a redress option involves a legal claim in a court or tribunal. Another means adopted here was not to visualise information about the comparative costs, rates of use and success rates of particular redress options in the mapping tool itself (but we did provide this information to policy makers in our reports). While case prevalence is useful for lawyers, judges and policy makers etc., it hides many nuances, which can mislead and bias people's decisions.

\begin{figure*}
    \centering
    \includegraphics[width=.9\textwidth]{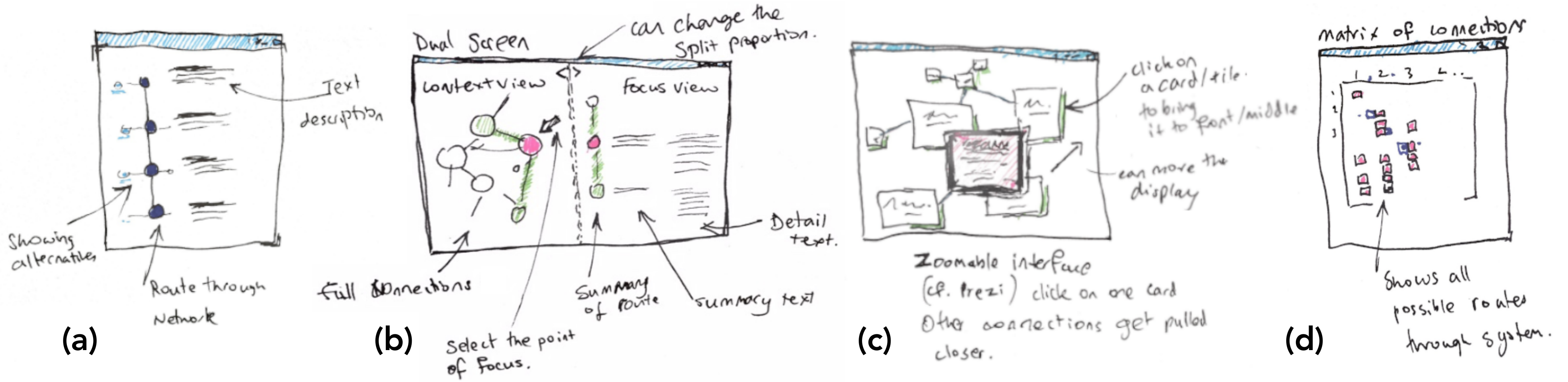}
    \vspace{-3mm}\caption{Four design sketches from our design exercise. a) Focus view, showing the main route and associated text, b) focus+context, a dual view with the full network on the left and specific path on the right. c) Zoomable interface with different cards that can include the example case study information. (d) Matrix view of full connections. Our chosen design follows ideas from a) and b).}\vspace{-4mm}
    \label{fig:designSketches}
\end{figure*}

Other valuable feedback from the workshops was that we should link clearly to where people can go to seek legal advice, this further reinforces that we are not providing legal advice ourselves, and crucially provides links at appropriate stages in the pathways to relevant providers of legal advice, advocacy, and other related support. Despite that we are not providing legal advice, and are linking to advice websites, workshop feedback noted that we do need to provide some information about timing, as certain legal claims must be made within three months of the decision under dispute. So, we do highlight where time limits apply and remind users of something called a `pre-action protocol' these protocols are largely there to ensure that people have explored all alternative options for redress before seeking legal action. 

The workshops also highlighted specific details, relevant to each case study. In education, school exclusions was chosen for mapping mainly because this was the issue about which most concern had been expressed in the initial education workshops due to the lack of independent scrutiny mechanisms for most exclusions. The system for challenging an exclusion does not present many choices for an individual, so the pathways are not particularly complex in terms of visualisation. Perhaps the most challenging issue was teasing out the law on the different types of exclusion, as the rights to challenge an exclusion vary depending on whether it is a permanent or fixed-term exclusion, and if the latter on how long the exclusion is for. 

\subsection{Design of the Explanatory tool -- Artemus}
\label{sec:DesignOfTheExplnatoryTool}

\begin{figure}[h]
    \centering
    \includegraphics[width=\columnwidth]{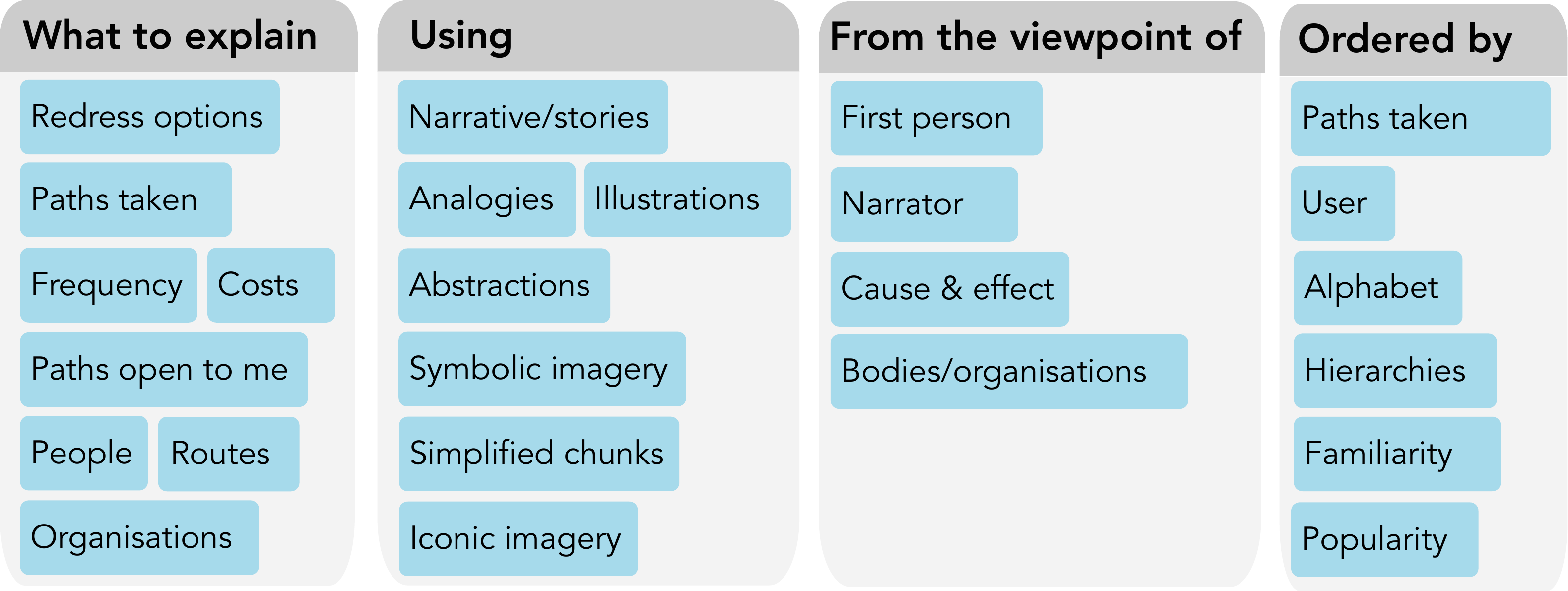}
    \caption{When designing the explanatory visualisation we kept coming back to these four questions: what are we explaining? What are we using to explain it? We are explaining it from what viewpoint? It is organised and ordered by what method?}
    \label{fig:WhatUsingViewpointOrdered}
    \vspace{-3mm}
\end{figure}

\begin{figure*}
    \centering
    \includegraphics[width=\textwidth]{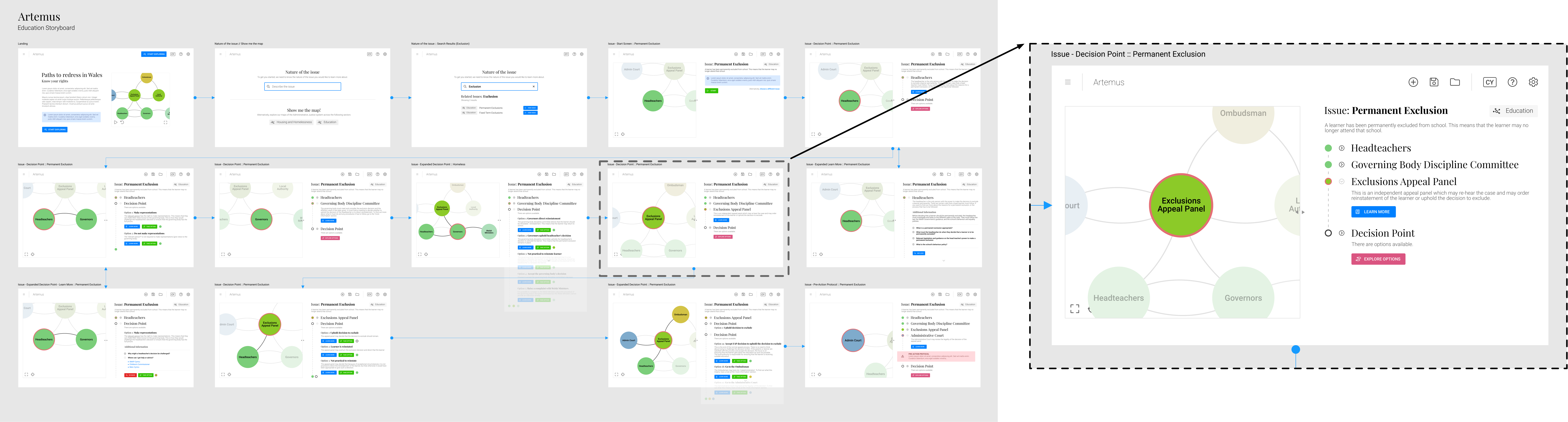}
    \caption{Storyboard of administrative justice redress paths in education. Developed for the second workshop in phase \texorpdfstring{\phaseTWOitem}{2}, showing an overview of all storyboard panels and zoom into the issue of someone being permanently excluded from a School. All tiles in the overview, except the first three, were screenshots from the prototype implementation.}\vspace{-3mm}
    \label{fig:designStoryboard}
\end{figure*}

Co-designing the explanatory visualisation meant that the design process started at the beginning of the project and continued throughout. We performed a focused design study during Phase~2 and the start of Phase~3. We followed the process from the Five Design-Sheet (FdS)~\cite{RobertsHeadleandRitsos16_FDS_TVCG} thinking and sketching design solutions (see~\autoref{fig:designSketches}); ideating potential solutions, refining to few designs, honing the designs and generating wire-frames that we shared at the second workshop (see~\autoref{fig:designStoryboard}). We considered other strategies (cf.\ \cite{Sed2012a,McKennaETAL2014,Munzner2009}), and especially the learning objectives framework of Adar and Lee~\cite{AdarLee2021_communicativeVis}, but the visual emphasis of the FdS, with user-embedded, iterative low-fidelity co-design processes, matched better to our explanatory goals. Explanatory visualisations need to present information in a correct and clear way. Consequently through the design process we needed to keep validating ideas against the goals and requirements of the user (\autoref{sec:UserTask}). It is not only a matter of merely generating new ideas, or alternative design solutions, but to make sure that the ideas are appropriate. Through this process we kept coming back to four questions: (1) What are we explaining? (2) What methods or strategies are we using to explain the ideas? (3) The output that we are creating is being displayed from which viewpoint? (4) How is the output organised or ordered? These questions provide a useful checklist to develop explanatory visualisations and are summarised in~\autoref{fig:WhatUsingViewpointOrdered}.

\textbf{What are we explaining?} In Phase~2 we revisit this question (from our first presentation of these issues in \autoref{sec:UserTask}). In fact, through the project we continuously reflected on this question. There are many concepts we could explain: possible paths, path popularity, success rates, redress cost in money, or focus on organisations and their relationships. However the workshop attendees confirmed our focus on redress routes, and we did not want to distract from our core message by displaying other information. For example, displaying cost or frequency may bias viewpoints and could encourage or dissuade people from taking one route instead of another.

\textbf{What strategies are we using to explain?}
Interacting with the stakeholders allowed us to develop the schematic diagrams (\autoref{fig:HomelessDraft}
and \autoref{fig:HomelessDraftImproved}), understand user requirements (\autoref{sec:UserTask}) and refine, confirm and reject different design ideas. Not only did these schematics help to define redress paths and options, but helped to clarify what was important (or not) in the design idea. For instance, during the workshops people commented on ``the complexity of the redress routes in the schematics'' and suggested the ``routes could be simplified into sub-parts''. Oversimplification could bias understanding, but simplification does help people focus on specific detail. Consequently, we sketched several examples to simplify the paths (see~\autoref{fig:designSketches}a). They suggested that it could ``show the whole system, yet allow people to learn about specific routes'', and that ``different bodies involved should be clear; perhaps by different colours for local authorities, courts, tribunals and so on''. We sketched a dual-view system, see~\autoref{fig:designSketches}b.
Furthermore, participants discussed whether colour should reflect the type of body (court, tribunal, etc.), reflect the stage in the redress pathway (redress one, redress two, etc.), or represent if the redress body was a devolved Welsh institution (the Ombudsman and some tribunals) or non-devolved (the courts). Their comments and feedback helped influence our final design. For instance, we chose colour to reflect the type of body, displayed simplified routes, and allow users to choose what to view (following \autoref{fig:designSketches}b).

Another way to consider this challenge is in terms of semiotics~\cite{Bertin1983} (iconic, indexical or symbolic sign types). This idea can be expressed by asking: `what is our vehicle of explanation'? Perhaps the visual explanation uses a physical resemblance to represent the signified. For example, we could use photographs of a School Governor or Ombudsmen to help explain the people involved. Or we could imply specific ideas; where a smiley face could present a good decision to make or show photographs of different types of journeys to explain the types of road ahead if one redress path were taken in comparison to another. Or use symbols to help explain parts of the process. For instance we could create different symbols to represent an appeal, judicial review or complaint. Another possibility could be to make vignettes from the examples, which could be placed on `cards'~\cite{Strobelt_ETALDocumentCards_2009}, as sketched in \autoref{fig:designSketches}c. When people `request more information' a larger card could appear, which would include more detail. Another possibility was to demonstrate connections in a matrix view (\autoref{fig:designSketches}d). But, the stakeholders considered that `cards' may not be clear, and a matrix view was too abstract and would not be understood by the users.

While visualisation and animation are often used to help explain concepts~\cite{RobertsETAL2018_EVF} we considered other strategies, but through discussion at the workshops we rejected these ideas. For example, we could use a narrator to speak simplified descriptions, or use illustrations with text. Analogies can be used to help explain things. Analogies help share similar concepts of abstraction, perhaps between information that someone knows and what they do not know. We could compare the Ombudsmen to an honest parent, or present an advocate as an unbiased friend. But sometimes analogies are misunderstood, confuse people and subtle differences can mislead. Analogies, however, are often used in teaching, where for instance, the educator questions the student about their knowledge and relates new subjects with content that they know already.  While we could ask people about their knowledge of the judicial system (in a quiz, for instance) we decided that this was too intrusive and difficult to generalise.

\textbf{Display the output from which viewpoint?} In~\autoref{sec:UserTask} we have already discussed this question, and decided on the first person view. However, during our design process we re-considered different options:  first person, third person, cause and effect, comparison view or from the point of view of the bodies and organisations involved. For instance, it could be possible to explain redress options through videos that show people meeting an advocate, walking to the court room for a judicial review, and so on. Such direct symbology can help people understand what would be required if they choose a particular redress route, but may be time consuming to create and observe. Through the first person viewpoint we would be able to add in specific examples, with named characters, which would help users relate and reference the example situations.  For instance, Sara may be having a problem with her child at School and has been sent a letter expelling her child. What are her options? Sara's first option would be to contact the School.

\textbf{How is the output organised?} In this question we considered both the visual layout and also the conceptual order of this information. Displaying all the nodes at once (as per the schematic diagrams that we used to understand different options; \autoref{fig:HomelessDraft} and \autoref{fig:HomelessDraftImproved}) may be useful for the expert, but maybe less useful for the public. The information on the network map needs to be displayed in context, with authentic information and real-world examples.
We could display some of the paths, but then which paths, and how would they be displayed? One solution could be to display a sub-path as a strip, starting the node with one `of interest' from the user. Strip maps are geographical maps that follow a route, which could be rivers, roads pipelines and so on. This is an \textit{explanatory journey}. We decided to focus on the strip map idea (\autoref{fig:designSketches}a,b), because it was understandable and encouraged by the workshop participants. However, one challenge is deciding where a user should start. We discussed different strategies, such as showing the whole map, giving a simplified network, and other orderings, such as cost, popularity of redress routes or alphabetical, but decided to use a user-selected entry point. The person types text into a text-field that the system interprets to locate the appropriate starting point. E.g., someone may type ``I have just been made homeless'' and the program would then start with the ``Local Authorities'' and homelessness node.

\section{Phase \texorpdfstring{\phaseTHREEitem}{3} -- Tool implementation, Reflection}
\label{sec:phase3}
\begin{figure}[H]
    \centering
    \includegraphics[width=\columnwidth]{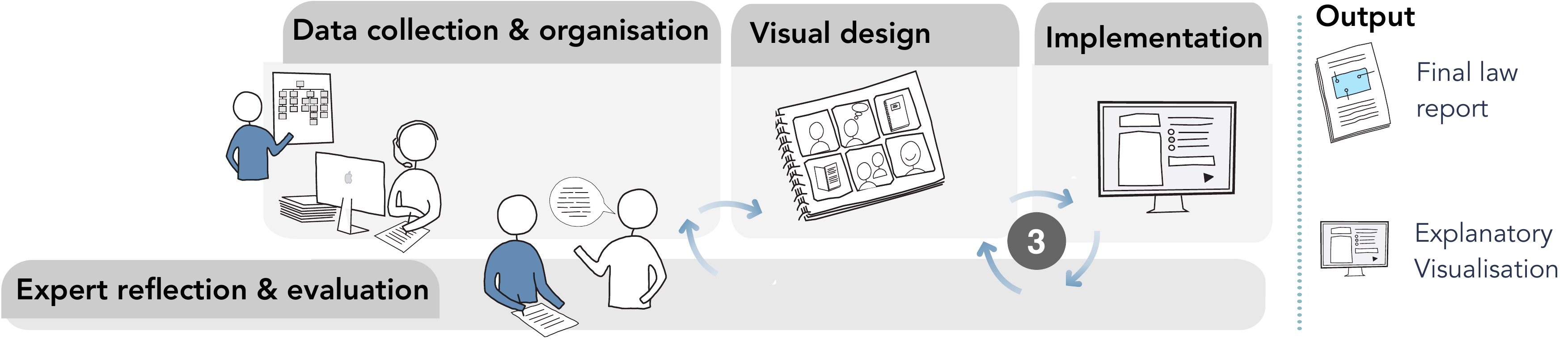}
    \vspace{-5mm}
    \caption{In Phase~3 we developed the final explanatory tools, reflecting on the design with expert feedback, and delivered the final law brief.}
    \label{fig:FlowPhase3}
\end{figure}

In Phase~3 (\autoref{fig:FlowPhase3}) we held the second set of workshops (using the same structure as the first set) and discussed the designs and draft implementation using the storyboard (\autoref{fig:designStoryboard}). We created both Welsh and English storyboards. At the time of the workshop the implementation was partially working, consequently we used some wire-framed mock-ups (the first three panels of the storyboard) alongside screenshots of our prototype (the remaining panels). We wanted to demonstrate an early version of the prototype, because it would allow us to make changes to it with feedback from workshop participants. After the workshops we continued to refine the written examples, finalised the explanatory visualisation tool, and delivered the final report. Bilingual development was important, due to the dual status of the English and Welsh languages in the administrative justice system in Wales. Drafting the content bilingually meant that we simultaneously delivered Welsh and English versions. In fact, we presented the Welsh version of Artemus at the Welsh National Eisteddfod in August 2019.

\subsection{Implementation -- Artemus explanatory visualisation for administrative justice}
We developed Artemus using cross-platform web technologies. We used  CytoscapeJS~\cite{10.1093/bioinformatics/btv557} for the maps, ReactJS and Redux, a modern toolset for creating Web-based user interfaces, and the Cloud Firestore real-time database. We map nodes and edges to system processes with JSON.
\autoref{fig:teaser} demonstrates Artemus on the homelessness use-case. The node-link diagram has directional edges between authority bodies, which is directly taken from the law briefing document. 
We deliberately designed a minimalist user interface, which was refined through feedback from experts and workshop participants.
The network visualisation gives an overview of possible routes. People can zoom out to see the full network, and the current pathway is shown in dark lines with potential routes elided by changing the colour of the edges to grey. People can gain further details-on-demand by selecting `learn more', which opens up more text.
Users are presented with choices on the right view, which records their journey as folded blocks of information. In \autoref{fig:teaser}D the user is offered 12 options, that can be viewed by scrolling the window.
As the user progresses through decision points, the folded blocks of previous decisions begin to build up a user's pathway. Previous steps can be revisited by opening the folded blocks, and any decision can be changed to explore alternative pathways.

\subsection{Expert reflection and evaluation (second workshops)}
Feedback from the workshop participants was very positive. 
By this stage, many experts thought that the tool would be  useful in the hands of those in the advice sector, and used by individuals to gain a sense of how to frame a grievance in a way that it could be pursued within the administrative justice system. It became clear that the advisers at our workshops imagined using Artemus, sitting with an individual and walking them through possible routes of redress. Bodies, without the remit to deal with complaints, are frequently approached by the public for help and they considered that they could use the tool to signpost individuals to relevant assistance. In one of the workshops, where we had advanced-level students attend (aged 16 and above), for them seeking information in this way through applications is the norm (these were well educated young adults). Likewise, the parents of children with special educational needs who attended our education law and redress workshops were well-educated professional people, used to seeking information online and would not have difficulty using Artemus themselves.
In the area of education, a solicitor considered that the tool would ``redress the knowledge imbalance between the professionals and members of the public'' and allow the latter a better chance of enforcing their rights. The tool was considered valuable in ``clarifying how the different cogs of the Welsh administrative justice system fit together''. It was also considered that, depending on its development, it might enable bodies such as the Welsh Commissioners (for Children, Older People, the Welsh Language, and Future Generations) to identify areas where there might be systemic problems. A government official considered that the tool had ``considerable potential'' for use within the Welsh Government Civil Service, including for ``policy development, stakeholder engagement and staff training'', and featured on the Welsh Parliament's Senedd research news~\cite{NasonGov2020}.

\section{Summary and reflections}
We have developed Artemus: an explanatory tool to visualise and explain administrative justice paths of redress, focusing on administrative law in housing and homelessness, and education. Not only have we designed and developed the bilingual explanatory tool, but we have delivered several in-depth reports on these issues in administrative law~\cite{nason2020public,NasonGov2020}; advice that is being used by stakeholders, lawyers and Government officials. We used a collaborative co-design method, which allowed us time to reflect and gain ongoing feedback from stakeholders. With Artemus, people start at a redress point of interest and build a personal journey by choosing routes of redress. Summary information is shown and more detailed explanations can be expanded on request. There is little guidance in the research domain over best practices for creating explanatory visualisations. But, methods and processes that we have followed can act as a blueprint for others to follow.

\newcommand*{\img}[1]{%
    \raisebox{-.3\baselineskip}{%
        \includegraphics[
        height=\baselineskip,
        width=\baselineskip,
        keepaspectratio,
        ]{#1}~~~%
    }%
}

\newcommand\myspacing{1mm}
\newcommand\myspacenewline{1.5mm}

\noindent\begin{minipage}{\columnwidth}
\lettrine[lines=3]{{\includegraphics[width=8mm]{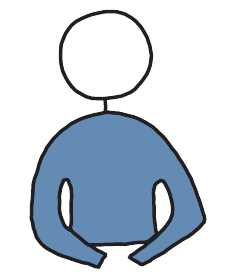}}}{\hspace{\myspacing}}
\textbf{Understand the user.} We asked who will be using the explanatory visualisation and what concepts will we explain. Why, where and when they would use it. 
\end{minipage}\vspace{\myspacenewline}
%====================
\noindent\begin{minipage}{\columnwidth}
\lettrine[lines=3]{{\raisebox{1mm}{\includegraphics[width=8mm]{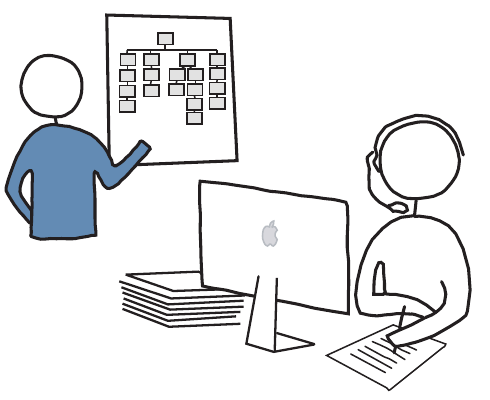}}}}{\hspace{\myspacing}}
\textbf{Collect, understand and organise data.} We researched, organised information, output reports, developed example scenarios and co-created draft designs (\autoref{fig:HomelessDraft} and \ref{fig:HomelessDraftImproved}). 
\end{minipage}\vspace{\myspacenewline}
%====================
\noindent\begin{minipage}{\columnwidth}
\lettrine[lines=3]{{\raisebox{1mm}{\includegraphics[width=8mm]{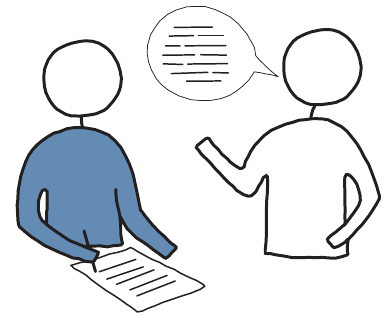}}}}{\hspace{\myspacing}}
\textbf{Validate the information gathered.} We held bilingual workshops to validate, correct and develop additional details of the redress paths. 

\end{minipage}\vspace{\myspacenewline}
%====================
\noindent\begin{minipage}{\columnwidth}
\lettrine[lines=3]{{\raisebox{1mm}{\includegraphics[width=8mm]{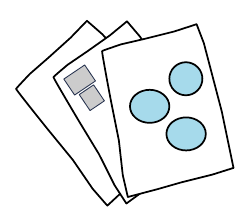}}}}{\hspace{\myspacing}}
\textbf{Design your solution.} Using co-design, developing from draft network diagrams (\autoref{fig:HomelessDraft} and \autoref{fig:HomelessDraftImproved}), and using the structure from the Five Design-Sheet~\cite{RobertsHeadleandRitsos16_FDS_TVCG} we sketched different designs. 
\end{minipage}\vspace{\myspacenewline}
%====================
\noindent\begin{minipage}{\columnwidth}
\lettrine[lines=5]{{\includegraphics[width=8mm]{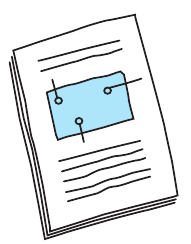}}}{\hspace{\myspacing}}
\textbf{Implement and keep validating the solution, share it.} We created the explanatory visualisation, shared it with experts at the workshops and improved the visualisation based on their feedback. We published the visualisation along with the written work, to be used by advisers, experts and the public.\hspace{6mm}
\end{minipage}
%====================

Our research demonstrates that there are many complex ideas in law and administrative justice, that people would like to understand but the information is not easily accessible, in one place or held in a way that is easy to comprehend. Explanatory visualisation has the power to make this complex information accessible. However, because there is no single database, before a developer can create the visualisation they need to collect, understand and organise the data and scenarios. We examined the legislation, case law, previous research, statistics and reports from governments, tribunals and so on. We compiled a register of institutions, looked at general law on dispute resolution and wrote an account of the different areas of weaknesses, disputes and parties involved. This represented a huge quantity of work, but was necessary to understand possible redress paths. This data-gathering challenge occurs across law and administrative justice. While we manually collected this data,  we believe that this data extraction could be aided through the use of learning algorithms, which is an important area of future work.

Working collaboratively and co-designing with experts (from a wide range of bodies and organisations, from judges, lawyers, ombudsmen) and those aggrieved, has been a rewarding and important process. While the final design reflects ideas of the original schematic diagrams used to discern the redress paths, through considering the different approaches (see~\autoref{sec:DesignOfTheExplnatoryTool}) the experts favoured the network design. One said ``it is similar to how my mind envisions the redress paths''. We have many examples of instances where ideas have come from this process that would probably not have surfaced. For instance, the ``ok to do nothing'' option came from a workshop participant; who after presenting their case realised that ``it is too easy to get caught up in a process you don't know when to stop''. Parents can become so accustomed to fighting battles for their children's education and do not know when it wise not to pursue a particular redress. Another observation from the second workshop was about the `on demand approach'. One participant said ``it forced me to consider the essential details before investigating any additional information'' and went on to say ``sounds basic but it can be too easy to display visualisations which seem helpful, but people drown in the information''. 
Especially raised in the Independent Advice Providers Forum were issues of how to cater for other languages, aggrieved or vulnerable individuals, or users with various disabilities. Some members of the general public could understand Artemus to a certain degree without an adviser walking them through it, whereas others may find the visual representation especially valuable, but would need an adviser to help them fully understand it. This is well recognised in the move to online courts where there are provisions for `assisted digital' for those that request it whereas many people navigate the online forms and systems on their own.

What is perhaps distinctive and quite central to the nature of the administrative justice system itself is that people have choices and sometimes overlapping routes to redress, narrowing down to one single optimal legal answer is not reflective of the nature of the subject. Hence why the design allows people to explore the different pathways, and to understand the consequences both of the choices that they themselves can make, and also the consequences of the different outcomes (decisions) that particular bodies like tribunals and courts might make. The other key benefit is having the context view as well as the pathway, as this helps users understand the relationship between the pathway and the broader system. Explanatory systems should help people understand ideas (e.g., complex legal issues), and decide how they want to explore and learn the information.

Meeting with and hearing from lawyers, judges, ombudsmen and so on, has helped us understand the different issues, and hear their views on visualisation, and there is strong support for transparency and clarity of presentation of concepts. Indeed, our workshop participants expressed the importance of  visualisation and clarity of explanation in administrative justice.  Where visualisations of legal processes are used, they are designed for professionals and not general advisers, and are not interactive. The split screen on Artemus, use of focus+context, and strip-map design, worked well. People could view their journey and the bigger picture at the same time. The visualisation also helped to show the complexity of the subject. It allowed experts to observe gaps and inconsistencies in the redress process, and has already led to reforms within the judicial process. While breaking the redress routes into ``nodes'' and ``edges'' does represent a different way for the experts to think about administrative justice, it allowed them to see complexity and coherence in the redress systems, and observe structures that would not otherwise have been spotted or appreciated. 

\acknowledgments{
The authors wish to thank Dr Huw Pritchard (Cardiff University), Dr Helen Taylor (Cardiff Metropolitan University) for their input and expert advice. This work was supported by the \href{https://www.nuffieldfoundation.org/project/paths-to-administrative-justice-in-wales}{Nuffield Foundation} grant JUS43523. We acknowledge \href{www.userfocus.co.uk/uxstencil}{Userfocus} for the UX people used in the diagrams. }

\bibliographystyle{abbrv-doi}
\bibliography{ms}
\end{document}